\documentclass[11pt,a4paper]{article}
\usepackage[truedimen,margin=33mm]{geometry}

\usepackage{mathrsfs}
\usepackage{amssymb}
\usepackage{color}
\usepackage{amsmath}
\usepackage{ascmac}
\usepackage{amsthm}
\usepackage{booktabs}
\usepackage{setspace}
\usepackage{multirow}

\usepackage{titlesec}
\titleformat*{\section}{\large\bfseries}
\titleformat*{\subsection}{\it}

\usepackage[pdftex]{graphicx}
\usepackage[pdftex]{color}

\def\be{{\beta}}

\def\ep{{\varepsilon}}
\def\la{{\lambda}}

\def\om{{\omega}}
\def\th{{\theta}}

\def\bbe{{\text{\boldmath $\beta$}}}

\def\bphi{{\text{\boldmath $\phi$}}}

\def\beh{{\widehat \be}}

\def\muh{{\widehat \mu}}

\def\muh{{\widehat \mu}}

\def\nuh{{\widehat \nu}}

\def\mut{{\widetilde \mu}}

\def\bphih{{\widehat \bphi}}

\def\0{{\text{\boldmath $0$}}}

\def\u{{\text{\boldmath $u$}}}

\def\x{{\text{\boldmath $x$}}}
\def\y{{\text{\boldmath $y$}}}

\def\E{{\text{\boldmath $E$}}}

\def\Var{{\rm Var}}

\def\CV{{\rm CV}}

\def\E{{\rm E}}

\usepackage{color}

\title{{\bf Small Area Estimation with Spatially Varying Natural Exponential Families}}
\date{}

\begin{document}
\doublespacing
\maketitle

\vspace{-2cm}
\begin{center}
Shonosuke Sugasawa$^1$, Yuki Kawakubo$^2$ and Kota Ogasawara$^3$
\end{center}

\noindent
$^1$Center for Spatial Information Science, The University of Tokyo\\
$^2$Graduate School of Social Sciences, Chiba University\\
$^3$Department of Industrial Engineering, School of Engineering, Tokyo Institute of Technology

\vspace{0.5cm}
\begin{center}
{\large\bf Abstract}
\end{center}
Two-stage hierarchical models have been widely used in small area estimation to produce indirect estimates of areal means.
When the areas are treated exchangeably and the model parameters are assumed to be the same over all areas, we might lose the efficiency in the presence of spatial heterogeneity.
To overcome this problem, we consider a two-stage area-level model based on natural exponential family with spatially varying model parameters. 
We employ geographically weighted regression approach to estimating the varying parameters and suggest a new empirical Bayes estimator of the areal mean.
We also discuss some related problems, including the mean squared error estimation, benchmarked estimation, and estimation in non-sampled areas.
The performance of the proposed method is evaluated through simulations and applications to two data sets.


\newpage

\section{Introduction}
\label{sec:int}
Small area estimation is widely used to produce reliable estimates of areal means with small, or even zero, sample sizes.
When area-specific sample sizes are small, it is well recognized that the direct estimator based only on area-specific samples has high variability and is not appropriate for practical use.
Hence, we need to ``borrow strength" from related areas and produce indirect (model-based) estimates of areal means.
To this end, two-stage hierarchical models have been widely used as standard statistical tools in small area estimation.
For comprehensive reviews of small area estimation techniques, see Pferffermann (2013) and Rao and Molina (2015).

We suppose we are interested in the true areal mean $\th_i$ for $i=1,\ldots,m$, where $m$ is the number of areas.
Let $y_i$ be the direct estimator of $\th_i$ based only on the available samples within the $i$th area, so that $y_i$ is typically unstable, i.e., the coefficient of variation $\sqrt{\Var(y_i|\th_i)}/y_i$ is unacceptably large.
In order to improve the accuracy of $y_i$ by ``borrowing strength" from information on related areas, hierarchical models have been widely adopted.
The most common model in small area estimation is a two-stage hierarchical normal model known as the Fay--Herriot (FH) model (Fay and Herriot, 1979), which is applicable for continuous values.
For more general cases such as count or binary, models based on a natural exponential family with conjugate priors (Ghosh and Maiti, 2004) or generalized linear mixed models (McCulloch and Searle, 2001; Jiang, 2006) are available.
Although such conventional methods do not take account of geographical information, there is a growing body of literature that develops effective methods with use of geographical information. 
There are mainly two types of the way of adopting geographical information.
One is the use of spatial correlated random effects as considered in Bandyopadhyay et al. (2009), Marhuenda et al. (2013), Pratesi and Salvati (2009), Schmid et al. (2016) and Wakefield (2007) among others, and the other is based on geographically weighted regression (Brundson et al., 1996; Fotheringham et al., 2002) as considered in Chambers et al. (2014), Chandra et al. (2012, 2015, 2017) and Salvati et al. (2012) among others.

Although generalized linear mixed models are widely used for non-normal data, it is well-known that fitting generalized linear mixed models would be computationally intensive due to the intractable integral appeared in the marginal likelihood function.
Hence, incorporating spatially correlated random effects or geographical weighted regression into generalized linear mixed models would be computationally burdensome as well, so that it would not be user-friendly.  
To overcome this difficulty and extend the body of knowledge on this topic, we here focus on models based on the natural exponential family with quadratic variance function proposed in Ghosh and Maiti (2004) and incorporate spatially varying parameters, which is a similar structure used in geographically weighted regression, into the model.
We employ the local likelihood method (Tibshirani and Hastie, 1987) to estimate varying parameters, and the bandwidth in the local likelihood is selected via cross validation.
The main advantage of the proposed model is the analytical tractability, that is, the marginal likelihood as well as the Bayes estimator in the proposed model can be obtained in a analytical way unlike generalized linear mixed models.
Hence, the (local) maximum likelihood estimator and empirical Bayes estimator can be easily computed.

The rest of the paper is organized as follows.
In Section \ref{sec:main}, we propose spatially varying empirical Bayes methods and discuss some related problems, including mean squared error (MSE) estimation, benchmarked estimation, and estimation in non-sampled areas.
In Section \ref{sec:simu}, we evaluate the finite sample performance of the proposed methods through simulations.
In Section \ref{sec:app}, we show the results of two applications, the first to Scottish lip cancer data using the Poisson--gamma model and the second to Spanish poverty rate data using the binomial--beta model.
Finally, Section \ref{sec:conc} discusses the results and concludes the paper.

\section{Small area models with spatially varying natural exponential families}
\label{sec:main}


\subsection{Spatially varying models and local likelihood estimation}\label{sec:SVM}
Let $m$ be the number of areas; $\{y_i,\x_i\}_{i=1,\ldots,m}$ be the sampled data, where $y_i$ is the direct estimator of an area mean $\mu_i$, satisfying $\E[y_i|\mu_i]=\mu_i$; and $\x_i$ be a vector of covariates associated with $y_i$. 
Typically, $y_i$ is an unstable estimator of $\mu_i$ in the sense that $\Var(y_i|\mu_i)$ is large because of the small sample size within the area.
Ghosh and Maiti (2004) proposed the following hierarchical model based on natural exponential family:
\begin{equation}\label{model}
\begin{split}
f(y_i|\th_i)&=\exp\big\{n_i(\th_i y_i -\psi(\th_i))+c(y_i,\phi_i)\big\},\\
\pi(\th_i; \bphi)&=\exp\big\{\nu(m_i\th_i-\psi(\th_i))+C(\nu,m_i)\big\},
\end{split}
\end{equation}
where $m_i=\psi'(\x_i^t\bbe)$ with $\psi'(t)=d\psi/dt$ is the canonical link function, $\th_i$ is a natural parameter, $n_i$ is a known scalar (dispersion) parameter, $\bphi=(\bbe^t,\nu)^t$ is a vector of unknown model parameters common to all the areas, and $\psi(\cdot), c(\cdot,\cdot)$, and $C(\cdot,\cdot)$ are functions specific to each distribution.
For some typical applications, $n_i$ is equal to the sample size within area $i$.
However, in general, $n_i$ is just the dispersion parameter but not necessarily the area sample size.
For example, in the application to count data illustrated in Section \ref{sec:Scottish}, $n_i$ denotes the expected number of the cases of lip cancer in the $i$th area.
Under the model, the area mean $\mu_i$ is expressed as 
$$
\mu_i=\E[y_i|\th_i]=\psi'(\th_i),
$$
noting that $\E[\mu_i]=m_i$ under (\ref{model}).
Moreover, it is assumed that the conditional variance is a quadratic function of the conditional mean $\mu_i$, namely $\Var(y_i|\th_i)=n_i^{-1}Q(\mu_i)$, where $Q(x)=v_0+v_1x+v_2 x^2$ for known constants $v_0$, $v_1$, and $v_2$, which are not simultaneously zero.

In this paper, we introduce a spatially varying structures in the model (\ref{model}), that is, we allow the model parameters in (\ref{model}) to vary spatially. 
We propose the following model: 
\begin{equation}\label{spmodel}
\begin{split}
f(y_i|\th_i)&=\exp\big\{n_i(\th_i y_i -\psi(\th_i))+c(y_i,n_i)\big\},\\
\pi(\th_i; \bphi(\u_i))&=\exp\big\{\nu(\u_i)(m_i(\u_i)\th_i-\psi(\th_i))+C(\nu(\u_i),m_i(\u_i))\big\},
\end{split}
\end{equation}
where $m_i(\u_i)=\psi'(\x_i^t\bbe(\u_i))$, $\u_i=(u_{1i},u_{2i})$ represent the coordinates of the $i$th area, and $\bphi(\u_i)=(\bbe(\u_i)^t,\nu(\u_i))^t$ denote the spatially varying model parameters. 
Note that the first-stage model of $y_i|\th_i$ is the same as (\ref{model}), while the prior distributions of $\th_i$ are different over the areas.
Under the model (\ref{spmodel}), the Bayes estimator of $\mu_i$ under the quadratic loss is given by
$$
\mut_i\equiv \mut_i(y_i,\bphi(\u_i))=\frac{n_iy_i+\nu(\u_i)m_i(\u_i)}{n_i+\nu(\u_i)}.
$$
Let $\bphih(\u_i)$ be the estimator of $\bphi(\u_i)$ discussed below.
Then the empirical Bayes estimator of $\mu_i$ can be obtained as $\muh_i=\mut_i(y_i,\bphih(\u_i))$.

To estimate the spatially varying parameters $\bphi(\u_i)$, we adopt the local likelihood method (Tibshirani and Hastie, 1987) and estimate $\bphi(\u_i)$ by maximizing the following locally weighted log-likelihood function:
\begin{equation}\label{Q}
\ell(\bphi(\u_i))=\sum_{k=1}^m w_b(\|\u_i-\u_k\|) \log M(y_k;\bphi(\u_i)),
\end{equation}
where $w_b(\cdot)$ is a user-specified kernel function with bandwidth $b$ and
$$
M(y_k;\bphi(\u_i)) = \exp\left\{ C(\nu(\u_i),m_k(\u_i))-C(n_k+\nu(\u_i),\mut_k(y_k,\bphi(\u_i))) \right\},
$$
which is proportional to the marginal likelihood with $m_k(\u_i) = \psi'(\x_k^t\bbe(\u_i))$ and
$$
\mut_k(y_k, \bphi(\u_i)) = {n_ky_k + \nu(\u_i)m_k(\u_i) \over n_k + \nu(\u_i) }.
$$
Note that the above function has an analytical form since the function $C(\cdot,\cdot)$ is uniquely and analytically determined by the distribution of $\th_i$. 
A common choice of the kernel $w_b(\cdot)$ would be the Gaussian kernel defined as 
$$
w_b(\|\u_i-\u_k\|)=\exp\left(-\frac{\|\u_i-\u_k\|^2}{2b^2}\right),
$$
where $b$ is the bandwidth controlling the rate at which the weight declines depending on the distance between two locations.

In practice, the bandwidth $b$ is unknown and we need to adaptively specify the value.
To this end, we use the following cross-validation (CV) criterion based on the marginal likelihood:
\begin{equation}\label{cv}
\CV(b)=\sum_{i=1}^m \log M(y_i;\bphih_{(-i)}(\u_i; b)),
\end{equation}
where $\bphih_{(-i)}(\u_i; b)$ is the estimates of $\bphi(\u_i)$ based on (\ref{Q}) under bandwidth $b$ without using $y_i$.
The optimal $b$ is the maximizer of $\CV(b)$, which can be obtained by using numerical methods.
We used the golden section search (Brent et al., 1973) over the interval $[b_\ell,b_u]$, with positive $b_\ell$ and $b_u$, for example, $b_\ell=0.01$ and $b_u=2\max_{i,k}\|\u_i-\u_k\|^2$.
Finally, the overall estimation procedure is as follows:

\bigskip
\noindent
\textbf{Estimation Procedure}
\begin{itemize}

\item[1.]
Decide the bandwidth $b$ by CV criterion in \eqref{cv}.
\medskip

\item[2.]
Estimate spatially varying parameter $\bphi(\u_i)$ for each area $i$ by maximizing the locally weighted log-likelihood function \eqref{Q} with bandwidth value decided by step 1.
\medskip

\item[3.]
Calculate the empirical Bayes estimates $\muh_i = \mut_i( y_i, \bphih(\u_i) )$, where $\bphih(\u_i)$ is the estimates of spatially varying parameter obtained in step 2.

\end{itemize}


\subsection{Mean squared error estimation}\label{sec:SV-mse}
In real-life applications, it is important to measure the uncertainty of the empirical Bayes estimator in order to assess the reliability of the estimates.
Traditionally, an estimator of MSEs has been used for this purpose (see Prasad and Rao (1990) and Datta et al. (2005)).
The MSE of the empirical Bayes estimator $\muh_i$ can be expressed as  
\begin{align*}
{\rm MSE}_i=\E\left[(\muh_i-\mu_i)^2\right]&=\E\left[(\mut_i-\mu_i)^2\right]+\E\left[(\muh_i-\mut_i)^2\right]\\
&\equiv R_{1i}(\bphi(\u_i))+ R_{2i}(\bphi(\u_i)),
\end{align*}
since $\mut_i=\E[\mu_i|y_i]$.
The first term corresponds to MSE of the conditional mean $\mut_i$ given the unknown parameters, thereby we can evaluate $R_{1i}(\bphi(\u_i))$ in the same way as Ghosh and Maiti (2004).
Then, it follows that 
\begin{equation*}
R_{1i}(\bphi(\u_i))=\frac{\nu(\u_i) Q(m_i(\u_i))}{(n_i+\nu(\u_i))(\nu(\u_i)-v_2)},
\end{equation*}
On the other hand, based on the theory of local likelihood (Tibshirani and Hastie, 1987), the estimator $\bphih(\u_i)$ would converge to the true $\bphi(\u_i)$ as $m\to\infty$.
Hence, the difference between $\muh_i$ and $\mut_i$ gets negligible as $m\to\infty$, so that the second term $R_{2i}(\bphi(\u_i))$ is expected to vanish as $m\to\infty$.
Hence, we may define $\widehat{\text{MSE}}_i^N=R_{1i}(\bphih(\u_i))$ as the naive (primitive) estimator of the MSE, which would be consistent under $m\to\infty$.
Although we do not give its rigorous proof, this property will be investigated in simulation studies given in Section \ref{sec:sim-MSE}.
It is known that, if $m$ is not large, $R_{2i}$ is not necessarily negligible, and the naive estimator could underestimate the true MSE.
Moreover, the plug-in estimator $R_{1i}(\bphih(\u_i))$ is known to have a considerable bias.
Therefore, the use of bias-corrected estimator of MSE is a standard approach in the context of small area estimation.

To construct a bias-corrected MSE estimator, we adopt the hybrid bootstrap approach employed by Butar and Lahiri (2003).
Let $\{y_1^b,\ldots,y_m^b\}$ be the parametric bootstrap samples generated from model (\ref{spmodel}) with $\bphi(\u_i)=\bphih(\u_i)$, and define $\bphih^b(\u_i)$ as the estimator computed from the bootstrap samples.
Then, the hybrid bootstrap MSE estimator is given by
\begin{equation}\label{spMSE}
\begin{split}
\widehat{\text{MSE}}_i^B
&=2R_{1i}(\bphih(\u_i))-\frac1B\sum_{b=1}^BR_{1i}(\bphih^b(\u_i))\\
& \ \ \ \ +\frac1B\sum_{b=1}^B\left\{\mut_i(y_i^b,\bphih^b(\u_i))-\mut_i(y_i^b,\bphih(\u_i))\right\}^2.
\end{split}
\end{equation}
Note that the last term corresponds to the estimator of $R_{2i}$, and first two terms correspond to a bias-corrected estimator of $R_{1i}$.
Here, we use an additive form for bias correction while estimating $R_{1i}$, although several other forms have been proposed (e.g. Hall and Maiti, 2006).


\subsection{Benchmarked estimation}\label{sec:bench}
A (weighted) sum of empirical Bayes estimates is not necessarily equal to the corresponding direct estimates, which is not preferable for practitioners.
Moreover, the empirical Bayes approach sometimes produces over-shrunk estimates, which results in inaccurate estimates of small area means.
To avoid these problems, the benchmarked estimator (Datta et al., 2011; Bell et al., 2013) has been used as a standard tool in small area estimation.
Here, we consider the constraint $\sum_{i=1}^mc_i\muh_i=\sum_{i=1}^mc_iy_i$ with some known weight $c_i$ satisfying $\sum_{i=1}^mc_i=1$.
A typical choice is $c_i=n_i/\sum_{k=1}^mn_k$.
From Datta et al. (2011), the constrained empirical Bayes estimator $\muh_i^{C}$ that minimizes the squared error $\sum_{i=1}^m\E[(\muh_i^{C}-\mu_i)^2]$ has the form
\begin{equation}\label{CEB}
\muh_i^{C}=\muh_i+\om_i\sum_{k=1}^m c_k\left(y_k-\muh_k\right),
\end{equation}
with $\om_i=c_i/\sum_{k=1}^mc_k^2$. 
The weight $c_i$ often satisfies $\max_{1\leq i\leq m}c_i=O(m^{-1})$ like $c_i=n_i/\sum_{k=1}^mn_k$.
Then, the difference between $\muh_i^C$ and $\muh_i$ decreases as the number of areas $m$ increases; namely, the differences are negligible when $m$ is sufficiently large.

Since the benchmarked estimator increases the MSEs compared to the empirical Bayes estimator, we need to assess how large the excess MSE is.
Regarding this issue, Steorts and Ghosh (2013) and Kubokawa et al. (2014) investigated the MSE estimators of benchmarked empirical Bayes estimators in area-level models using analytical or numerical methods.
Following Kubokawa et al. (2014), we consider a bootstrap method for evaluating the excess MSE.
The excess MSE is expressed as 
\begin{align*}
{\rm EMSE}_i
&=\E\left[(\muh^C_i- \mu_i)^2\right]-\E\left[ (\muh_i-\mu_i)^2\right]\\
&=\E\left[(\muh^C_i- \muh_i)^2\right]+2\E\left[(\muh^C_i- \muh_i)(\muh_i- \mut_i)\right].
\end{align*}
Therefore, the parametric bootstrap procedure used in the previous section enables us to estimate the excess MSE:
\begin{equation}\label{exMSE}
\widehat{{\rm EMSE}}_i
=\frac1B\sum_{b=1}^B\left(\muh^{C,b}_i-\muh_i^b\right)^2
+\frac2B\sum_{b=1}^B\left(\muh^{C,b}_i-\muh_i^b\right)\left\{\muh^b_i-\mut_i(\y_i^b,\bphih(\u_i))\right\},
\end{equation}
where $\muh_i^b=\mut_i(y_i^b,\bphih^b(\u_i))$, and $\muh^{C,b}_i$ is the benchmarked estimator (\ref{CEB}) by replacing $y_i$ and $\muh_i$ with $y_i^b$ and $\muh_i^b$, respectively.

\subsection{Estimation in non-sampled areas}\label{sec:SV-nonsamp}
Real applications could include small areas with zero sample sizes.
Let $j$ be the index of a non-sampled area and assume that the covariate $\x_j$ is available.
We can define the estimator of $\mu_j$ under the spatially varying model (\ref{spmodel}) as 
\begin{equation}\label{est-ns}
\mut_j(\bbe(\u_j))=m_j(\u_j)=\psi'(\x_j^t\bbe(\u_j)),
\end{equation}
with known $\bbe(\u_j)$.
The estimator of $\bphi(\u_j)=(\bbe(\u_j)^t,\nu(\u_j))^t$ can be obtained by maximizing the following local likelihood function: 
$$
\ell(\bphi(\u_j))=\sum_{k=1}^m w_b(\|\u_j-\u_k\|) \log M(y_k;\bphi(\u_j)),
$$
thereby we can compute the empirical version of (\ref{est-ns}).

\subsection{Typical Models}\label{sec:exm}
We here provide three typical models included in the proposed model (\ref{spmodel}).

\medskip\noindent
{\bf (Fay--Herriot model)} \ \ \ 
When we assume that distributions of $y_i|\th_i$ and $\th_i$ are both normal, with $n_i=D_i^{-1}, \nu(\u_i)=A(\u_i)^{-1}$, $\psi(\theta_i)=\theta_i^2/2$, $v_1=v_2=0$, and $v_0=1$, model (\ref{spmodel}) corresponds to the Fay--Herriot model (Fay and Herriot, 1979) with spatially varying parameters, described as 
$$
y_i=\x_i^t\bbe(\u_i)+\sqrt{A(\u_i)}b_i+\sqrt{D_i}\ep_i, \ \ \ \ i=1,\ldots,m,
$$
where the $b_i$s and $\ep_i$s are mutually independent standard normal random variables, and the $D_i$s are known sampling variances.
Under the model, the marginal distribution of $y_i$ is also normal, $\textrm{N}(\x_i^t\bbe(\u_i),A(\u_i)+D_i)$; thus, the Fisher scoring algorithm is easily implemented for maximizing local likelihood (\ref{Q}).

\medskip\noindent
{\bf (Poisson--gamma model)} \ \ \ 
When the distributions of $z_i(\equiv n_iy_i)|\mu_i$ and $\mu_i\equiv \exp(\th_i)$ are assumed to be Poisson and gamma, respectively, with $\psi(\th_i)=\exp(\th_i)$, $v_0=v_2=0$, $v_1=1$, and $C(\nu, m) = \nu m\log\nu - \log\Gamma(\nu m)$, model (\ref{spmodel}) is expressed as 
\begin{equation}\label{PG}
z_i|\mu_i\sim{\rm Po}(n_i\mu_i) \quad \mu_i\sim \Gamma(\nu(\u_i) m_i(\u_i),\nu(\u_i)), \ \ \ \ \ i=1,\ldots,m,
\end{equation}
where $\mu_1,\ldots,\mu_m$ are mutually independent, Po$(\la)$ denotes the Poisson distribution with mean $\la$, and $\Gamma(a,b)$ denotes the gamma distribution with density
$$
f(x)=\frac{b^a}{\Gamma(a)}x^{a-1}\exp(-bx), \ \ \ \ x>0.
$$
The model corresponds to the Poisson--gamma model proposed by Clayton and Kaldor (1987) with spatially varying model parameters.
It is well-known that the marginal distribution of $z_i$ is a negative binomial distribution with probability function
$$
f_m(z_i,\bphi(\u_i))=\frac{\Gamma(z_i+\nu(\u_i) m_i(\u_i))}{\Gamma(z_i+1)\Gamma(\nu(\u_i) m_i(\u_i))}\left(\frac{n_i}{n_i+\nu(\u_i)}\right)^{z_i}\left(\frac{\nu(\u_i)}{n_i+\nu(\u_i)}\right)^{\nu(\u_i) m_i(\u_i)},
$$
and the local likelihood (\ref{Q}) is similar to the likelihood of the geographical weighted negative binomial regression model suggested by Silva and Rodrigues (2014). 
For maximizing the weighted likelihood (\ref{Q}), we simply employed \verb+optim+ function available in R language.

\medskip\noindent
{\bf (binomial--beta model)} \ \ \ 
When the distributions of $z_i(\equiv n_iy_i)|\mu_i$ and $\mu_i\equiv \text{logistic}(\th_i)$ with $\text{logistic}(x)=\exp(x)/(1+\exp(x))$ are assumed to be binomial and beta, respectively, with $\psi(\theta_i)=\log(1+\exp(\theta_i))$, $v_0=0, v_1=1$, $v_2=-1$, and $C(\nu,m) = -\log B(\nu m, \nu(1-m))$, where $B(\cdot,\cdot)$ denotes beta function, model (\ref{spmodel}) is expressed as 
$$
z_i|\mu_i\sim {\rm Bin}(n_i,\mu_i) \quad \mu_i\sim {\rm Beta}(\nu(\u_i)m_i(\u_i),\nu(\u_i)(1-m_i(\u_i))), \ \ \ \ i=1,\ldots,m,
$$
where Beta$(a,b)$ denotes the beta distribution with density
$$
f(x)=B(a,b)^{-1}x^{a-1}(1-x)^{b-1}, \ \ \ 0<x<1.
$$
This model can be regarded as the extension of the binomial--beta model used by Williams (1975) in terms of the spatially varying hyperparameters.
Under the model, the marginal probability function of $z_i$ can be obtained as
$$
f_m(z_i,\bphi(\u_i))
={n_i \choose z_i}\frac{B(z_i+\nu(\u_i)m_i(\u_i), n_i-z_i+\nu(\u_i)(1-m_i(\u_i)))}{B(\nu(\u_i)m_i(\u_i),\nu(\u_i)(1-m_i(\u_i)))},
$$
thereby, the weighted likelihood (\ref{Q}) can be maximized by simply adopting \verb+optim+ function available in R language.

\section{Simulation studies}
\label{sec:simu}

\subsection{Estimation error comparison in sampled areas}\label{sec:comp1}
We first investigate estimation errors of the proposed estimator with the traditional estimator in finite samples. 
We consider the Poisson--gamma model described in Section \ref{sec:exm}.
As the coordinates $\u_i=(u_{1i},u_{2i})$ for $i = 1,\dots,m$, we use Scottish lip cancer data used in Section \ref{sec:app}.
In the dataset, we have $m = 56$ areas.
Covariate $x_i$ is generated from the uniform distribution on $(-1,1)$, which is fixed through simulation run.
As the data generating process, we consider the Poisson observation $z_i | \mu_i \sim \textrm{Po}(n_i \mu_i)$ and for $\mu_i$ we consider the following three scenarios:
\begin{align*}
\textrm{(I)}& \ \mu_i \sim \Gamma( \nu(\u_i) m_i(\u_i), \nu(\u_i) ), \quad \nu(\u_i) = 40\exp( u_{1i} + u_{2i} - 1 ) \\
\textrm{(II)}& \ \mu_i \sim \Gamma( \nu m_i, \nu ), \quad m_i = \exp( 0.5 + 0.5x_i ), \quad \nu = 40 \\
\textrm{(III)}& \ \log \mu_i = \log m_i(\u_i) + b_i, \quad b_i \sim \textrm{N}(0,1),
\end{align*}
with $m_i(\u_i) = \exp \{ \be_0(\u_i) + \be_1(\u_i) x_i \}$, $\be_0(\u_i) = u_{1i} - u_{2i} - 1$ and $\be_1(\u_i) = \sqrt{ u_{1i}^2 + u_{2i}^2 }$ for scenarios (I) and (III).
In each scenario, we divide $m = 56$ areas into 7 groups and set different values of $n_i \in \{ 5, 10, 20, 30, 40, 60, 100 \}$ for different groups.

We apply four models to the simulated data, our proposed spatially varying Poisson--gamma (SVPG) model, SVPG with benchmarking explained in Section \ref{sec:bench} (SVPG-B), spatially constant Poisson--gamma (SCPG) model and the following Poisson regression with conditional autoregression (PCAR) model (e.g. Escaramis et al., 2008; Goicoa el al., 2012)
$$
z_i | \mu_i \sim \textrm{Po}(n_i \mu_i), \quad \log\mu_i = \be_0 + \be_1x_i + e_i,
$$
where $(e_1,\dots,e_m)$ follows a Gaussian distribution with conditional autoregressive dependencies.
Following Ugarte et al. (2014), we used the integrated nested Laplace approximation (Rue et al., 2009) for fitting the PCAR model.



Based on $R=1000$ simulation runs, we simulate the area-level MSEs, defined as 
\begin{equation}\label{SVEB:sim-mse}
{\rm MSE}_i=\frac1R\sum_{r=1}^R\left( \muh_i^{(r)}-\mu_i^{(r)}\right)^2,
\end{equation}
where $\muh_i^{(r)}$ and $\mu_i^{(r)}$ denote the estimated and true values respectively of $\mu_i$ in the $r$th simulation run.
Then, we average the area-level MSEs over the same groups of $n_i$ values.
To compare the results between four methods, we compute the ratios of averaged MSEs of SVPG, SVPG-B and PCAR methods over averaged MSE of SCPG method.
The results are shown in Table \ref{tab:PG-MSE}.
Under scenario (I), the proposed SVPG and SVPG-B methods outperform the SCPG method.
The improvement of MSE is large especially in the areas for small $n_i$ value.
The performance of SVPG and SVPG-B is also better than PCAR except for the group for $n_i = 30$.
On the other hand, it is natural for the SVPG and SVPG-B methods to be inefficient compared to the SCPG method under scenario (II) since the former ones use only local information for estimating the hyperparameters.
However, it should be pointed out that the difference between the SV and SC methods is quite small in scenario (II) compared to the amount of improvement in scenario (I).
Under the scenario (III), in which none of the four models are the true data generating process, SVPG and SVPG-B methods perform the best or the second best.
The performance of SVPG and SVPG-B is very similar under all the scenarios.
However, benchmarking method is still important for practical use when the model based EB estimates are published by the government to keep consistency of the published values.

Next, we consider the performance of binomial--beta model.
For the coordinates $\u_i$, the auxiliary variable $x_i$ and the known scale parameter $n_i$, we use the same values as the simulation for the Poisson observations.
As the data generating process, we consider the binomial observation $z_i | \mu_i \sim \textrm{Bin}(n_i, \mu_i)$ and for $\mu_i$ we consider the following three scenarios:
\begin{align*}
\textrm{(I)} & \ \mu_i \sim \textrm{Beta}( \nu(\u_i)m_i(\u_i), \nu(\u_i)( 1 - m_i(\u_i) ) ), \quad \nu(\u_i) = 40 \exp( u_{1i} + u_{2i} - 1 ), \\
\textrm{(II)} & \ \mu_i \sim \textrm{Beta}( \nu m_i, \nu( 1 - m_i ) ), \quad m_i = \textrm{logistic}(0.5 + 0.5x_i), \quad \nu = 40, \\
\textrm{(III)} & \ \mu_i = \textrm{logistic}( \beta_0(\u_i) + \beta_1(\u_i)x_i + b_i ), \quad b_i \sim \textrm{N}(0, 0.25),
\end{align*}
with $m_i(\u_i) = \textrm{logistic}( \beta_0(\u_i) + \beta_1(\u_i)x_i )$ for scenario (I) and $\be_0(\u_i) = u_{1i} - u_{2i} - 1$, $\be_1(\u_i) = \sqrt{ u_{1i}^2 + u_{2i}^2 }$ for scenarios (I) and (III).
We apply four models to the simulated data, our proposed spatially varying binomial--beta (SVBB) model, SVBB model with benchmarking (SVBB-B), spatially constant binomial--beta (SCBB) model and the following logistic regression with conditional autoregression (LCAR) model:
$$
z_i | \mu_i \sim \textrm{Bin}(n_i, \mu_i) \quad \mu_i = \textrm{logistic}( \be_0, \be_1x_i + e_i ),
$$
where $(e_1,\dots,e_m)$ follows a Gaussian distribution with conditional autoregressive dependencies.

In the same manner as the Poisson observations, we simulated the area-level MSEs and calculated the ratio of the averaged MSEs over the same groups of $n_i$ values.
The results are given in Table \ref{tab:BB-MSE}.
We can see that the proposed spatially varying methods work well for the binomial observations as well.

\subsection{Estimation error comparison in non-sampled areas}

Next, we investigate the estimation errors in non-sampled areas as discussed in Section \ref{sec:SV-nonsamp}.
Based on the simulated data in Section \ref{sec:comp1}, we omit one area from each group, so that we observe $m = 49$ areas and the last $k = 7$ areas are non-sampled.
For the Poisson observations, we compare three methods, SVPG, SCPG and PCAR based on the scenario (I) with $R = 1000$ replications.
To compare the results, we simulated the area-level MSEs for $k=7$ areas based on $R=1000$ replications and calculated the ratios of the MSEs of SVPG and PCAR methods over the MSEs of SCPG method.
The results are shown in Table \ref{tab:PG-nonsample}.
From the table, we can see that our proposed SVPG method outperforms SCPG method except for the area with $n_i = 5$.
It is noted that PCAR method performs better than SCPG method for 5 out of 7 areas though PCAR performs worse than SCPG under scenario (I) of the simulation in the previous subsection.

For the binomial observations, we compare three methods, SVBB, SCBB and LCAR based on the scenario (I) with $R = 1000$ replications.
In the same manner as Poisson observations, we calculated the ratios of the simulated MSEs of SVBB and LCAR methods over the simulated MSEs of SCBB method.
The results are given in Table \ref{tab:BB-nonsample}, which shows the similar tendency to the Poisson case.


\subsection{Finite sample performance of MSE estimators}\label{sec:sim-MSE}
Finally, we investigate the finite sample performance of the MSE estimators developed in Section \ref{sec:SV-mse}.
Like the previous studies, we consider both the Poisson--gamma and binomial--beta models with spatially varying parameters.
Both for Poisson and binomial observations, we consider scenario (I) explained in Section \ref{sec:comp1} as the data generating process.
The coordinates $\u_i=(u_{1i},u_{i2})$ were generated from the uniform distribution on $(0,1)\times (0,1)$, and covariates $x_i$, from the uniform distribution on $(-1,1)$.
For the number of the areas $m$, we consider three situations $m = 30, 50$ and $80$.
In each scenario, we divide $m$ areas into 5 groups and set different values of $n_i \in \{ 10, 15, 20, 25, 30 \}$ for the group patterns of $n_i$.

We first simulate the MSEs (\ref{SVEB:sim-mse}) based on $R=1000$ simulation runs, which are used as the true values of the MSE in each area.
For estimating these true MSEs, we used estimators given in Section \ref{sec:SV-mse}: the naive estimator $\widehat{\text{MSE}}_i^\textrm{N}$ and the bootstrap estimator $\widehat{\text{MSE}}_i^\textrm{B}$ with $B=200$ bootstrap samples.
Based on $S=100$ iterations, we calculate the percentage relative bias (RB) and coefficient of variation (CV), which are respectively defined as 
$$
{\rm RB}_i=\frac1S\sum_{s=1}^{S}\frac{\widehat{\rm MSE}_i^{(s)}-{\rm MSE}_i}{{\rm MSE}_i},  \ \ \ \text{and} \ \ \ \ 
{\rm CV}_i=\sqrt{\frac1S\sum_{s=1}^{S}\bigg(\frac{\widehat{\rm MSE}_i^{(s)}-{\rm MSE}_i}{{\rm MSE}_i}\bigg)^2},
$$
where $\widehat{\rm MSE}_i^{(s)}$ is the MSE estimate of the $i$th area in the $s$th iteration, and ${\rm MSE}_i$ is the true MSE value of the $i$th area.

In Table \ref{tab:PG-mseestimate} and \ref{tab:BB-mseestimate}, $\textrm{RB}_i$s and $\textrm{CV}_i$s averaged over the same groups are reported for Poisson--Gamma and binomial--beta model, respectively.
From the tables, it can be seen that the naive estimators of the MSE have severe negative bias especially for the case that the number of areas is small.
On the other hand, bootstrap estimators can correct the bias even when the number of the areas is small.
In addition, for $m=30$ and $m=50$ cases, bootstrap estimators have smaller CV than naive estimators.
For $m=80$ case, both methods are comparable in terms of CV since the bias of the naive estimators get smaller because of the large number of areas.
It is also observed that RB does not change very much depending on $n_i$ while CV tends to be smaller with larger $n_i$.


\section{Examples}\label{sec:app}


\subsection{Scottish lip cancer data}\label{sec:Scottish}
We first apply the proposed method to Scottish lip cancer data collected during the 6 years from 1975 to 1980 in each of the $m=56$ counties of Scotland. 
These data were also analyzed by Clayton and Kaldor (1987).
The observed and expected number of cases are available for each county, respectively denoted by $z_i$ and $n_i$.
Moreover, the proportion of the population employed in agriculture, fishing, or forestry is available for each county, leading us to use it as covariate $\text{AFF}_i$, following Wakefield (2007). 
For each area, $i=1,\ldots,m$, we apply the spatially varying Poisson--gamma model:
\begin{equation}\label{SVPG-scott}
z_i|\mu_i\sim \text{Po}(n_i\mu_i), \ \ \ \mu_i\sim \Gamma(\nu(\u_i)\exp(\beta_1(\u_i)+\beta_2(\u_i)\text{AFF}_i),\nu(\u_i)),
\end{equation}
where $\u_i=(u_{i1},u_{i2})$, and $u_{i1}$ and $u_{i2}$ are the standardized longitude and latitude, respectively.

\begin{figure}[htbp]
\begin{center}
\includegraphics[width=7cm]{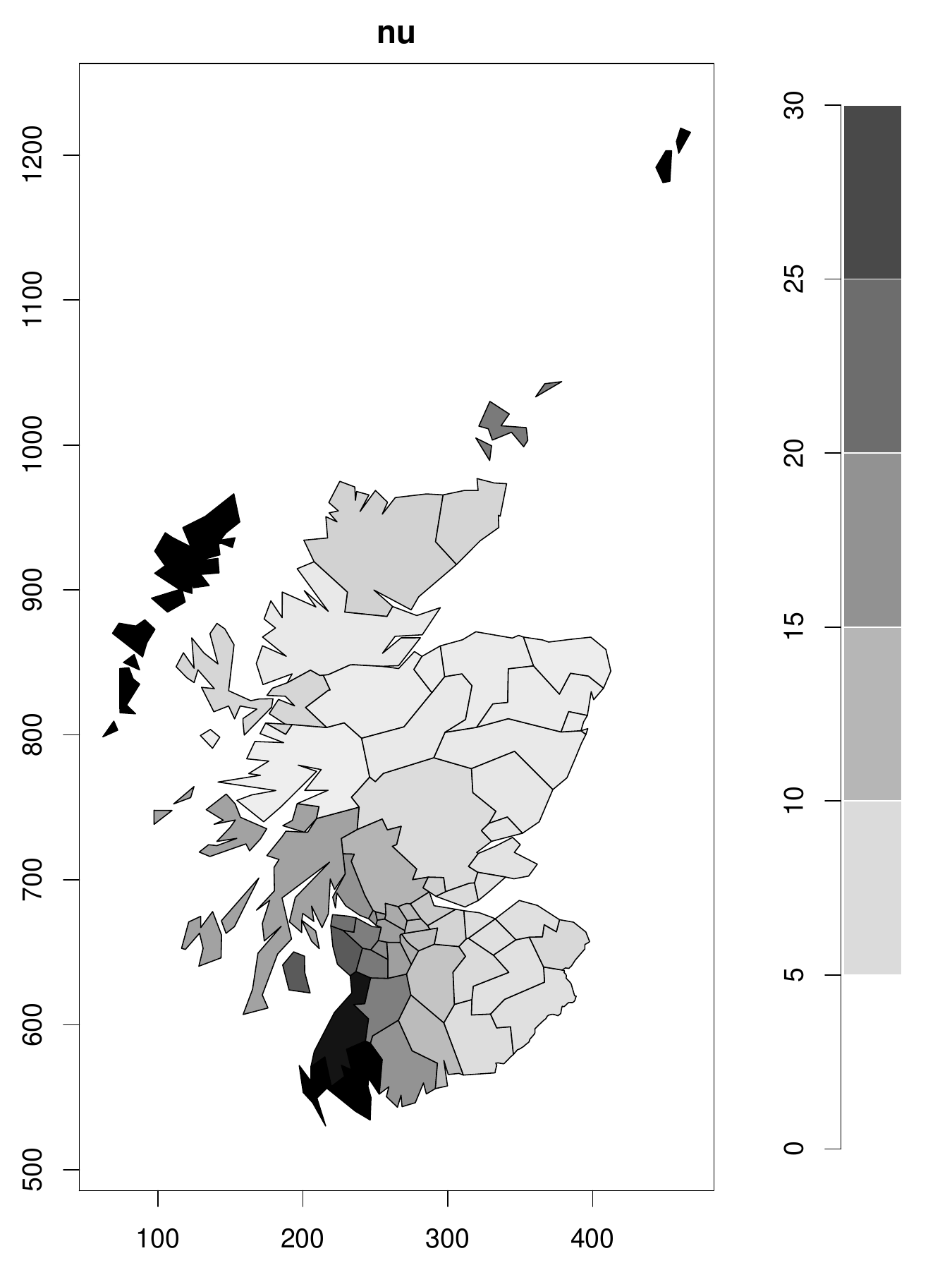}
\includegraphics[width=7cm]{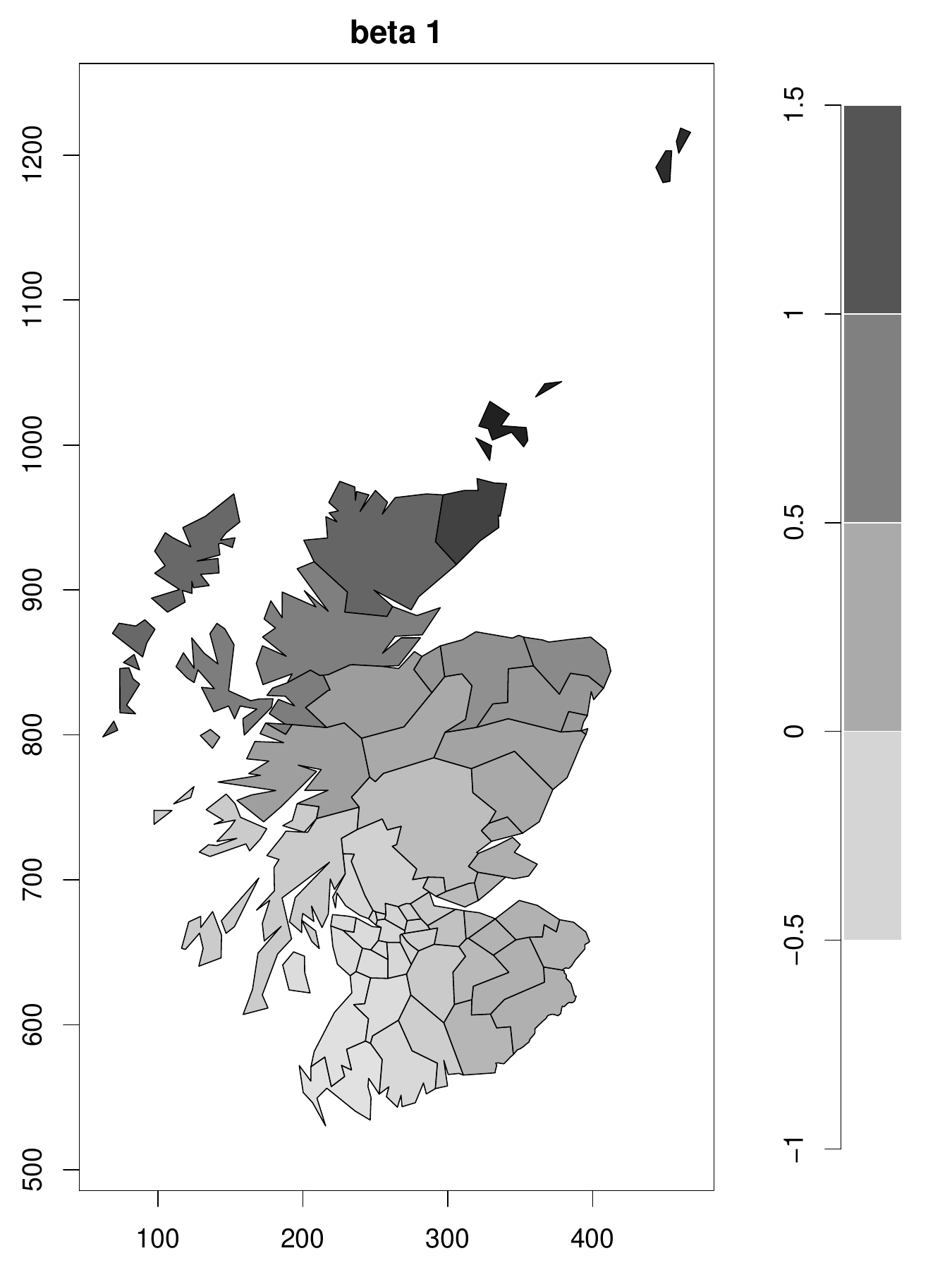}\\
\medskip
\includegraphics[width=7cm]{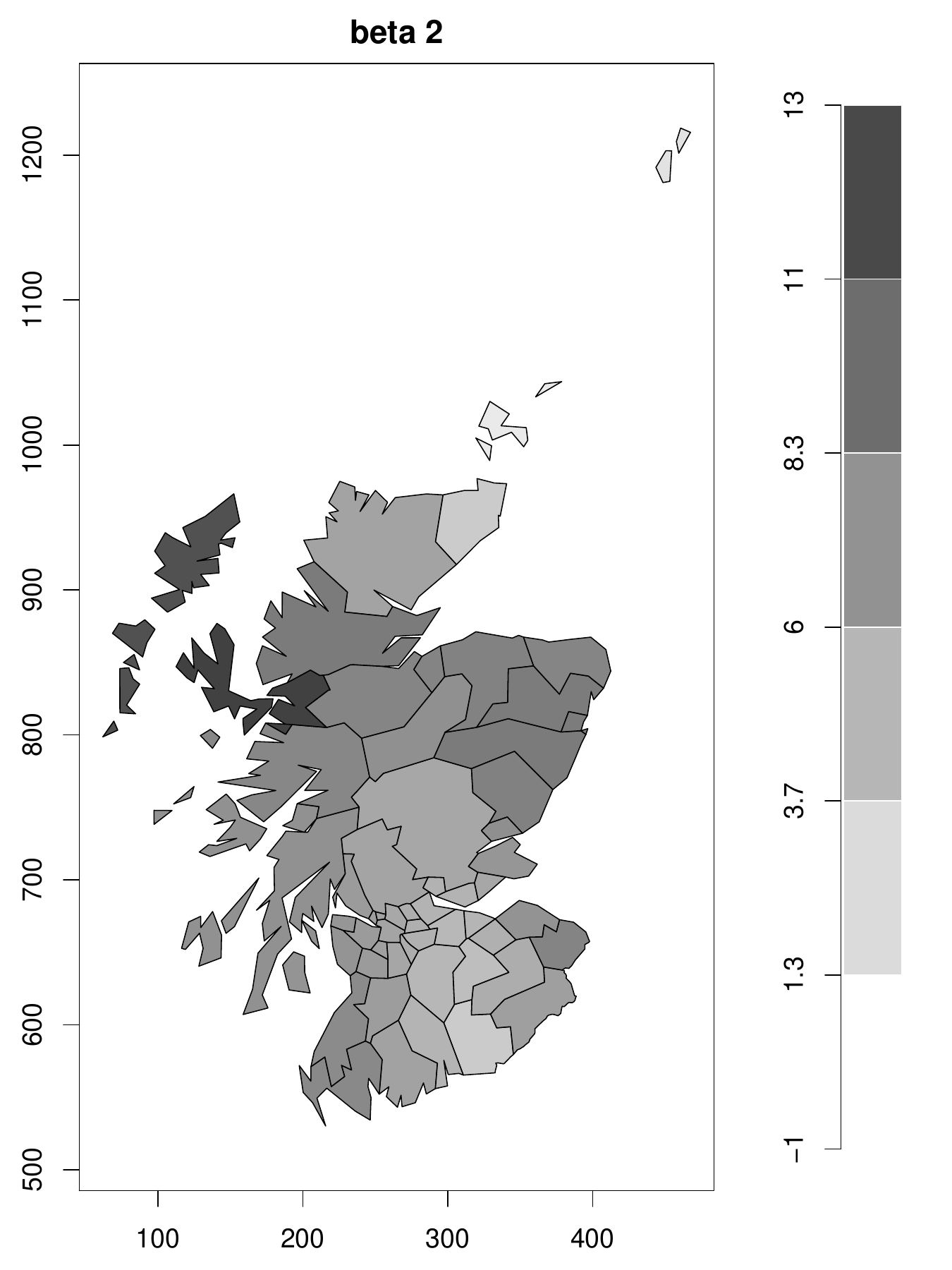}
\includegraphics[width=7cm]{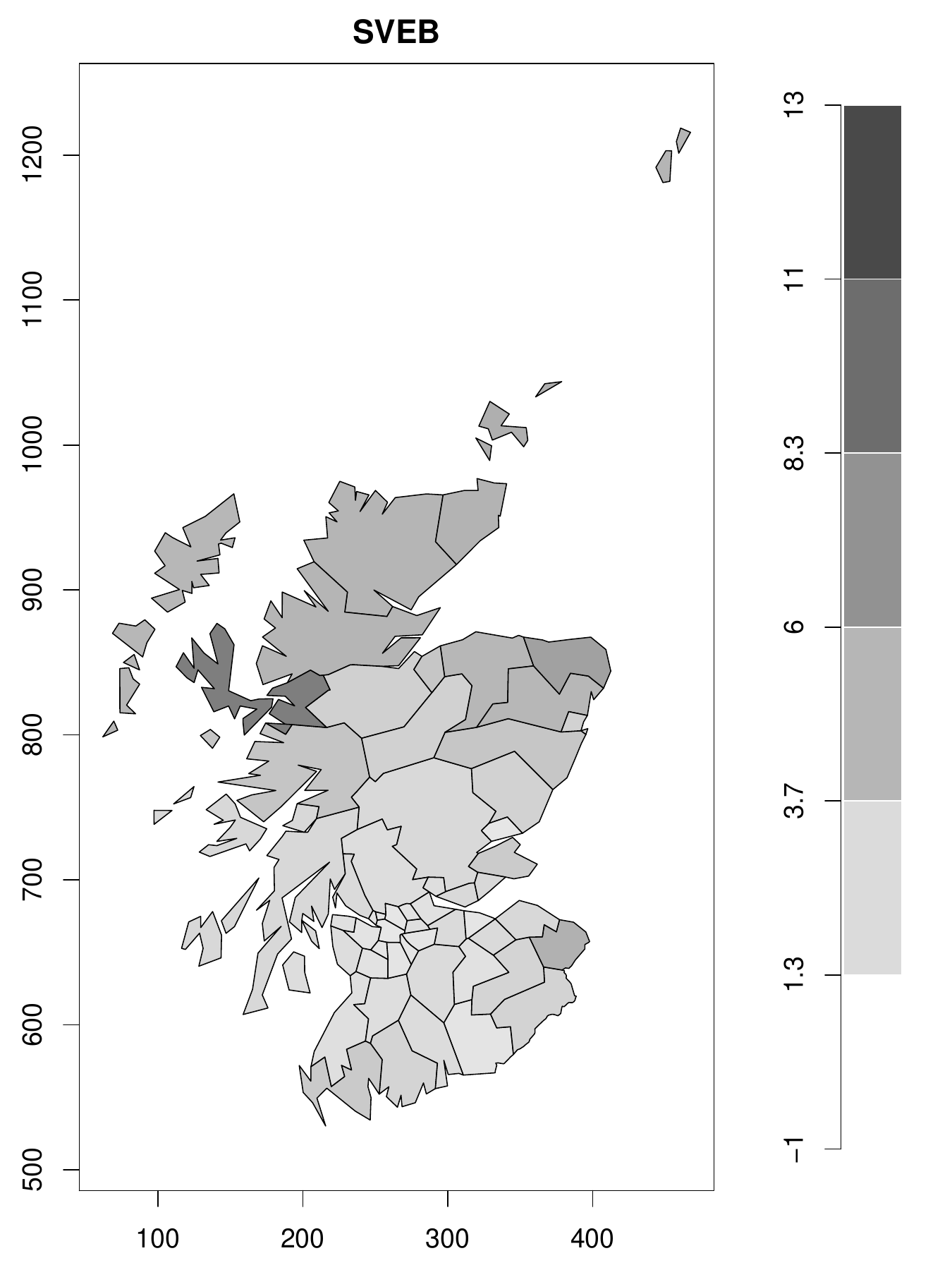}
\end{center}
\caption{Spatial distributions of $\nuh(\u_i)$ (top-left), $\beh_1(\u_i)$ (top-right), $\beh_2(\u_i)$ (bottom-left), and $\muh_i$ (bottom-right).
\label{fig:Scott-est}
}
\end{figure}

We first search for the optimal bandwidth by minimizing the criteria (\ref{cv}) and arrive at $b^{\ast}=0.900$.  
Then, we compute the estimates of the hyperparameters as well as the SVEB estimates of $\mu_i$ with $b=b^{\ast}$, which are shown in Figure \ref{fig:Scott-est}. 
According to Figure \ref{fig:Scott-est}, the hyperparameter estimates change dramatically from area to area.
For comparison, we apply the conventional Poisson--gamma model:
\begin{equation}\label{PG-scott}
z_i|\mu_i\sim \text{Po}(n_i\mu_i), \ \ \ \mu_i\sim \Gamma(\nu\exp(\beta_1+\beta_2\text{AFF}_i),\nu),
\end{equation}
and the maximum likelihood estimates of the hyperparameters are $\nuh=2.13, \beh_1=-0.15$, and $\beh_2=5.18$.

In order to investigated whether the parameters are spatially varying or not, we calculate the following statistics for each parameter: 
\begin{equation}\label{SV-stat}
\frac{\sum_{i=1}^mn_i\{\phi(\u_i)-\bar{\phi}\}^2}{\sum_{i=1}^mn_i},
\end{equation}
where $\phi\in\{\beta_0, \beta_1, \nu\}$ and $\bar{\phi}=\sum_{i=1}^mn_i\phi(\u_i)/\sum_{i=1}^mn_i$.
The $p$-value for testing the null hypothesis that there is no spatial variation can be numerically computed via the parametric bootstrap, where the bootstrap samples are generated from the model with spatially constant parameters, that is, (\ref{PG-scott}).
The $p$-values based on 1000 bootstrap samples are $0.096$, $0.795$ and $0.063$ for $\beta_0$, $\beta_1$ and $\nu$, respectively.
Hence, there could be spatial variations in $\beta_0$ and $\nu$ whereas there is not so strong evidence for spatial variation in $\beta_1$.

\begin{figure}[htbp]
\begin{center}
\includegraphics[width=15cm]{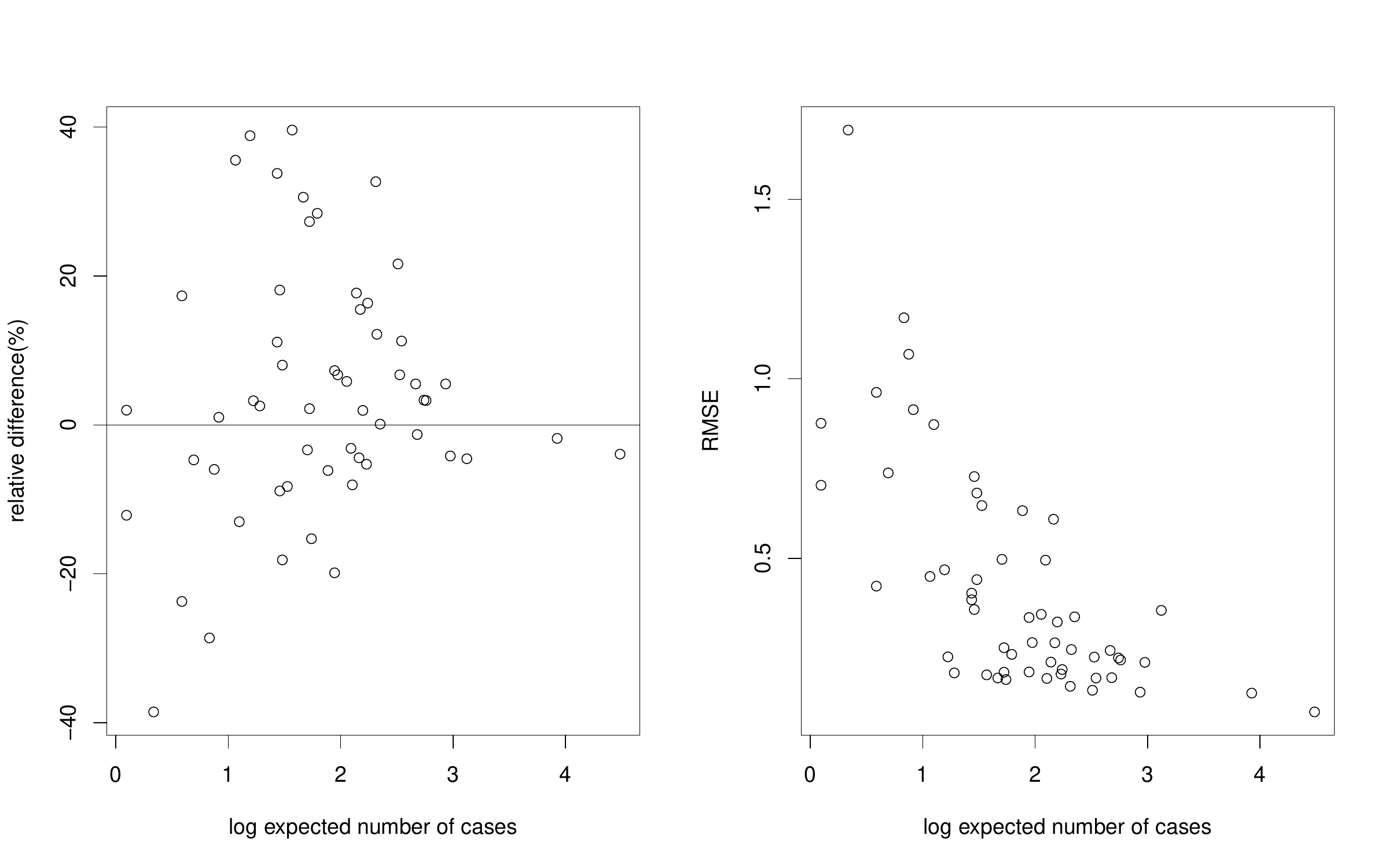}
\end{center}
\caption{Sample plots of the percentage relative difference between the SVEB and EB estimates: $100\times (\muh_i^{\rm EB}-\muh_i^{\rm SVEB})/\muh_i^{\rm SVEB}$ (left) and the squared root of MSE (RMSE) estimates of SVEB against $\log n_i$ (right) using Scottish lip cancer data.
}
\label{fig:Scott-EB}
\end{figure}

Let $\muh_i^{\rm SVEB}$ and $\muh_i^{\rm EB}$ be the SVEB estimates from (\ref{SVPG-scott}) and the empirical Bayes (EB) estimates from (\ref{PG-scott}), respectively.
In the left panel of Figure \ref{fig:Scott-EB}, we show the sample plot of the percentage relative difference $100\times (\muh_i^{\rm EB}-\muh_i^{\rm SVEB})/\muh_i^{\rm SVEB}$ against the log expected number of cases $\log n_i$.
We can observe that the differences are larger in areas with small $n_i$ and become smaller as $n_i$ increases.
This is because both the SVEB and the EB estimators are close to the direct estimator $y_i$ in areas with large $n_i$.
The right panel of Figure \ref{fig:Scott-EB} presents the sample plot of the square root of MSE (RMSE) estimates based on 500 bootstrap samples against $\log n_i$, revealing that the RMSE decreases as $n_i$ increases.

\begin{figure}[htbp]
\begin{center}
\includegraphics[width=15cm]{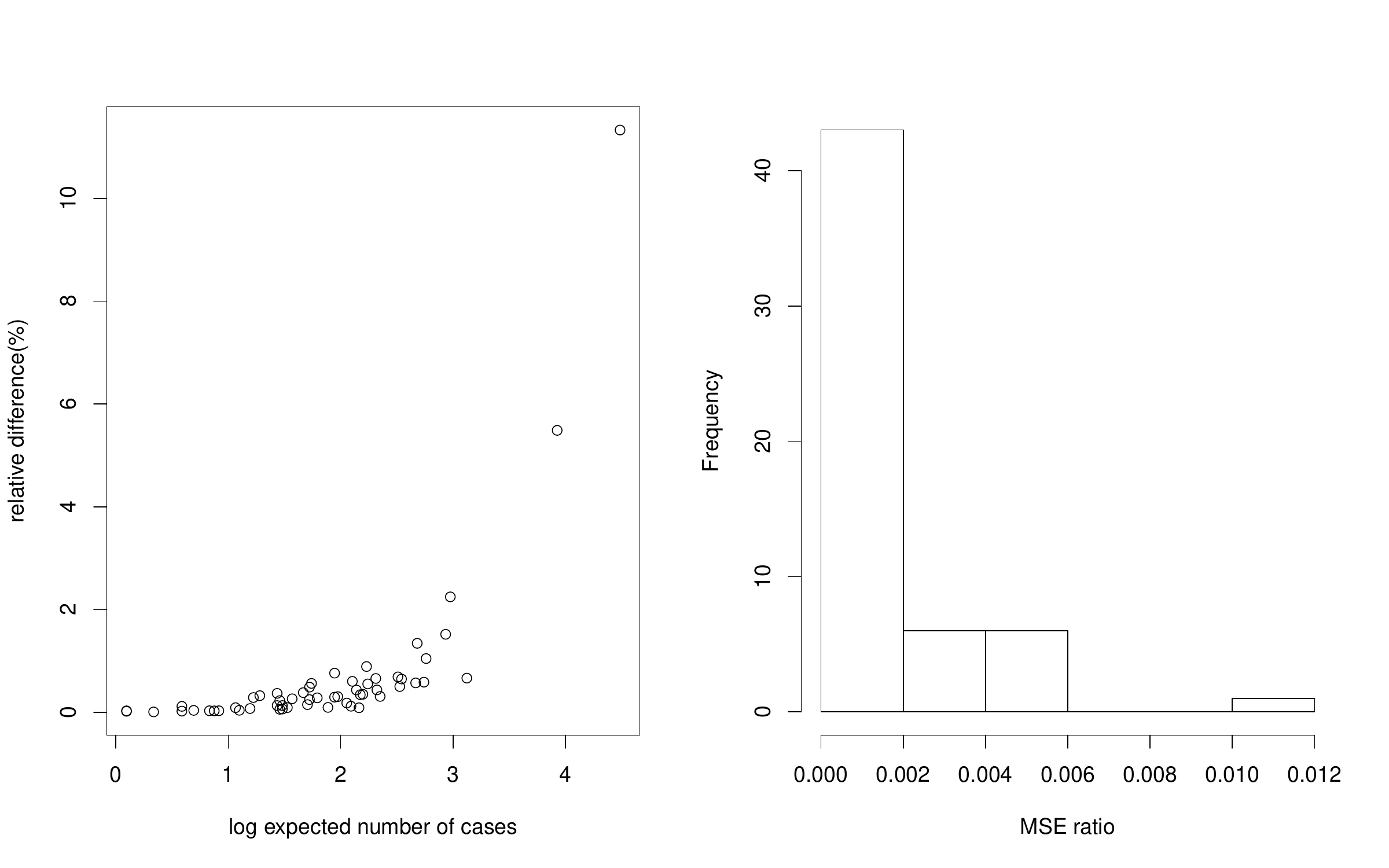}
\end{center}
\caption{Sample plots of the percentage relative difference between the SVEB and the benchmarked estimators: $100\times (\muh_i^{\rm C}-\muh_i)/\muh_i$ (left) and histogram of the excess MSE estimates (right) using Scottish lip cancer data.
\label{fig:Scott-CEB}
}
\end{figure}

We next compute the benchmarked estimator $\muh_i^{\rm C}$ from (\ref{CEB}) with the weight $c_i=n_i/\sum_{k=1}^mn_k$, and the relative differences to $\muh_i=\muh_i^{\rm SVEB}$ are presented in the left panel of Figure \ref{fig:Scott-CEB}.
The figure shows that the differences increase with respect to $n_i$ because of the choice of the benchmarking weight $c_i$. 
However, in most areas, the relative differences are smaller than $2\%$, so that $\muh_i^{\rm C}$ and $\muh_i$ are quite similar.
Based on $500$ bootstrap replications, we calculate the excess MSE estimates using (\ref{exMSE}) and compute the ratio to the MSE estimates of the SVEB.
The histogram of the ratio is given in the right panel of Figure \ref{fig:Scott-CEB}, which shows that the percentage of risk inflation is at most $1.4\%$.

Finally, we assess the performance of non-sampled area prediction.
We randomly omit 5 areas and predict $\mu_i$ in the omitted areas using SVEB and EB methods.
We then compute mean squared distance (MSD) between the predicted values and $y_i$ in the omitted areas.
We repeat this procedure for 100 times, and average values of MSD are 0.74 for SVEB and 1.22 for EB, so that SVEB would be more preferable in terms of non-sampled area prediction.


\subsection{Spanish poverty rate data}
Next, we use the synthetic income data set from Spanish provinces, which is available in R package \verb+sae+.
The data set contains unit data for $52$ areas.
Let $N_i, i=1,\ldots,m$ denote the population sizes of the areas.
Let $E_{ij}$ be the equivalized disposable income calculated following the standard procedure of the Spanish Statistical Institute, and $z$ be the poverty line.
The poverty rate for area $i$ is defined as $\mu_i=N_i^{-1}\sum_{j=1}^{N_i}I(E_{ij}<z)$.
Unfortunately, we do not observe all $E_{ij}$s but only observe $E_{ij}, j=1,\ldots,n_i$.
A direct estimator $y_i$ of $\mu_i$ is given by
$$
y_i=\frac{1}{n_i}\sum_{j=1}^{n_i}I(E_{ij}<z), 
$$ 
where we set $z$ as $0.6$ times the median of all the observed income $E_{ij}$s, following Molina and Rao (2010).
As area-level covariates, we use the area-level rates of females and labor, which are respectively denoted by $\text{fe}_i$ and $\text{lab}_i$.
Since two provinces, PalmasLas and Tenerife, are very far away from the other provinces, we omit their data in this study.
Then, we apply the following binomial--beta model for $i=1,\ldots,m$: 
\begin{equation}\label{SVBB-pov}
y_i|\mu_i\sim \text{Bin}(n_i,p_i), \ \ \ \mu_i\sim \text{Beta}(\nu(\u_i)m_i(\u_i),\nu(\u_i)(1-m_i(\u_i))), 
\end{equation}
where $m_i(\u_i)=\text{logistic}(\beta_1(\u_i)+\beta_2(\u_i)\text{fe}_i+\beta_3(\u_i)\text{lab}_i)$ and $\u_i=(u_{i1},u_{i2})$, and $u_{i1}$ and $u_{i2}$ are the standardized longitude and latitude, respectively.
For comparison, we also apply the conventional binomial--beta model: 
\begin{equation}\label{BB-pov}
y_i|\mu_i\sim \text{Bin}(n_i,p_i), \ \ \ \mu_i\sim \text{Beta}(\nu m_i,\nu(1-m_i)),
\end{equation}
with $m_i=\text{logistic}(\beta_1+\beta_2\text{fe}_i+\beta_3\text{lab}_i)$.

We find that the optimal bandwidth is $b^{\ast}=2.42$. Some empirical quintiles of the hyperparameter estimates are provided in Table \ref{tab:BB-pov}.
Table \ref{tab:BB-pov} shows that the median of spatially varying hyperparameter estimates in the spatially varying model (\ref{SVBB-pov}) is close to the point estimates in the conventional model (\ref{BB-pov}). 
We also employed the statistics (\ref{SV-stat}) for testing spatial variation, and calculated $p$-values based on 1000 parametric bootstrap samples, which are also reported in \ref{tab:BB-pov}. 
The $p$-values show that there would be spatial variations in regression coefficients, but there is not so strong evidence for spatial variation in $\nu$.

\begin{figure}[htbp]
\begin{center}
\includegraphics[width=15cm]{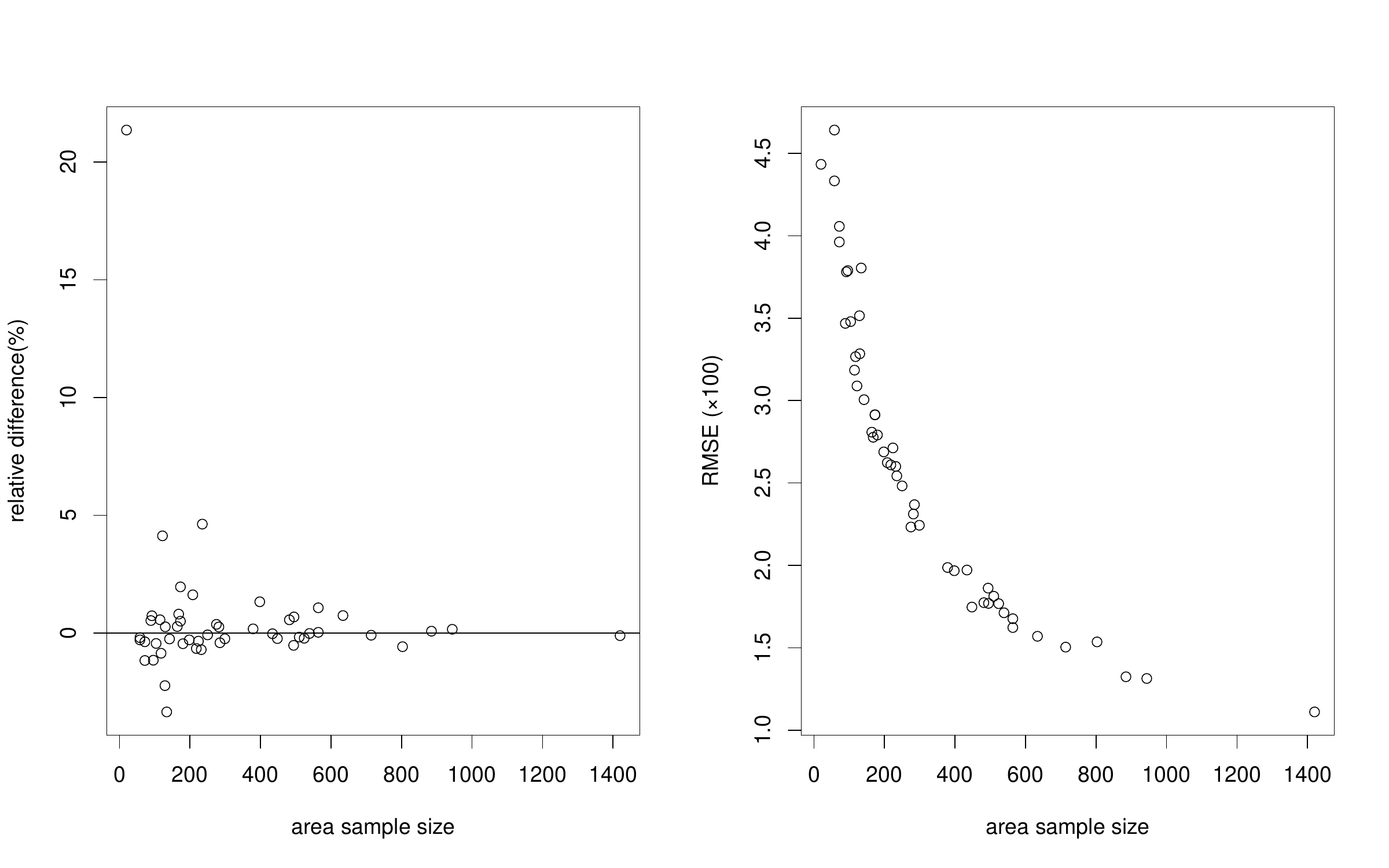}
\end{center}
\caption{Sample plots of the percentage relative difference between the SVEB and EB estimates: $100\times (\muh_i^{\rm EB}-\muh_i^{\rm SVEB})/\muh_i^{\rm SVEB}$ (left) and the squared root of MSE (RMSE) estimates of SVEB against $n_i$ (right) using Spanish poverty rate data.
\label{fig:BB-pov}
}
\end{figure}

The left panel in Figure \ref{fig:BB-pov} presents the percentage relative difference $100\times (\muh_i^{\rm EB}-\muh_i^{\rm SVEB})/\muh_i^{\rm SVEB}$, where $\muh_i^{\rm SVEB}$ and $\muh_i^{\rm EB}$ are the empirical Bayes estimates from (\ref{SVBB-pov}) and (\ref{BB-pov}), respectively.
We can observe that the differences are smaller than $6\%$ in all the areas except one, and they vanish as the area sample size $n_i$ increases.
Based on 500 bootstrap replications, we compute the MSE estimates of $\muh_i^{\rm SVEB}$. The RMSE estimates are given in the right panel of Figure \ref{fig:BB-pov}, showing the natural result that the MSE decreases with respect to $n_i$.
We next compute the benchmarked estimator of $p_i$ in model (\ref{SVBB-pov}), and we find that the percentage relative difference between the SVEB and benchmarked estimates are smaller than $0.15\%$, and the excess risks in benchmarking based on $500$ bootstrap samples are negligibly small.

Finally, we assess the performance of non-sampled area prediction.
We randomly omit 5 provinces and predict $p_i$ in the omitted areas using SVEB and EB methods.
We then compute mean squared distance (MSD) between the predicted values and $y_i$ in the omitted areas.
We repeat this procedure for 100 times, and average values of MSD multiplied 100 are 0.55 for SVEB and 0.58 for EB, so that SVEB is slightly better than BB in terms of non-sampled area prediction.

\section{Conclusions and discussion}\label{sec:conc}
We have developed SVEB methods based on the local likelihood estimation, in which the optimal bandwidth in a kernel function is determined by cross validation.
The model we considered can be regarded as a generalization of the two-stage hierarchical area-level models based on a natural exponential family, proposed by Ghosh and Maiti (2004).
The model includes the Fay--Herriot model, Poisson--gamma model, and binomial--beta model as special cases, so that it is applicable for continuous, count, and binary data.
We considered some problems, including the MSE and benchmarking estimations, as well as estimating in non-sampled areas.
The proposed methods were compared with the conventional non-spatial models through simulation and empirical studies. We found that the proposed method works well and improves the estimation accuracy of the traditional methods.

A possible drawback of the proposed method is its computational costs when the number of areas $m$ is large. 
For a specified bandwidth $b$, it requires $m$ times maximization of the weighted log-marginal likelihood (\ref{Q}) to compute the hyperparameter estimates in each area, thereby increasing the computational cost linearly depending on $m$.
A possible solution is to assume that $m$ areas can be classified in $G$ groups, where $G$ is much smaller than $m$, and that the hyperparameters remain the same in all the areas within the same group.
This can reduce the number of maximizations from $m$ to $G$ for each $b$.
However, the question remains as to how we may divide the areas efficiently.
However, a detailed consideration about this issue exceeds the scope of this paper, and we leave the problem to a future study.


\newpage

\begin{table}[htbp]
\begin{center}
\caption{The ratios of the area-level MSEs of SVPG, SVPG-B and PCAR methods over those of SCPG method. The values are averaged over the groups with the same $n_i$ values.}
\medskip
\begin{tabular}{llrrrrrrr}
\hline
& \multicolumn{1}{r}{$n_i$} & 5 & 10 & 20 & 30 & 40 & 60 & 100\\
\hline
& SVPG & 0.880 & 0.882 & 0.925 & 0.847 & 0.959 & 0.938 & 0.992\\
Scenario (I) & SVPG-B & 0.880 & 0.882 & 0.924 & 0.847 & 0.958 & 0.937 & 0.993\\
& PCAR & 1.107 & 0.986 & 0.973 & 0.794 & 1.045 & 1.042 & 1.087\\
\hline
& SVPG & 1.010 & 1.025 & 1.017 & 1.026 & 1.024 & 1.021 & 1.020\\
Scenario (II) & SVPG-B & 1.010 & 1.025 & 1.017 & 1.026 & 1.024 & 1.019 & 1.019\\
& PCAR & 1.150 & 1.800 & 1.277 & 1.369 & 1.457 & 1.592 & 1.599\\
\hline
& SVPG & 0.959 & 0.976 & 1.000 & 0.993 & 1.001 & 1.002 & 0.998\\
Scenario (III) & SVPG-B & 0.959 & 0.976 & 0.999 & 0.993 & 1.001 & 1.001 & 0.998\\
& PCAR & 0.912 & 0.967 & 1.011 & 1.006 & 1.000 & 1.009 & 1.002\\
\hline
\end{tabular}
\label{tab:PG-MSE}
\end{center}
\end{table}

\begin{table}[htbp]
\begin{center}
\caption{The ratios of the area-level MSEs of SVBB, SVBB-B and LCAR methods over those of SCBB method. The values are averaged over the groups with the same $n_i$ values.}
\medskip
\begin{tabular}{llrrrrrrr}
\hline
& \multicolumn{1}{r}{$n_i$} & 5 & 10 & 20 & 30 & 40 & 60 & 100\\
\hline
& SVBB & 0.961 & 0.947 & 0.961 & 0.926 & 0.991 & 0.965 & 1.002\\
Scenario (I) & SVBB-B & 0.961 & 0.947 & 0.961 & 0.926 & 0.990 & 0.963 & 1.004\\
& LCAR & 1.127 & 1.311 & 1.115 & 1.060 & 1.258 & 1.273 & 1.306\\
\hline
& SVBB & 1.013 & 1.017 & 1.013 & 1.012 & 1.010 & 1.005 & 1.007\\
Scenario (II) & SVBB-B & 1.013 & 1.017 & 1.012 & 1.012 & 1.010 & 1.005 & 1.007\\
& LCAR & 1.152 & 1.575 & 1.302 & 1.515 & 1.494 & 1.721 & 1.818\\
\hline
& SVBB & 0.976 & 0.969 & 0.984 & 0.971 & 1.001 & 0.995 & 1.004\\
Scenario (III) & SVBB-B & 0.976 & 0.969 & 0.983 & 0.971 & 1.001 & 0.993 & 1.006\\
& LCAR & 1.134 & 1.220 & 1.099 & 1.075 & 1.187 & 1.149 & 1.147\\
\hline
\end{tabular}
\label{tab:BB-MSE}
\end{center}
\end{table}

\begin{table}[htbp]
\begin{center}
\caption{The ratios of the MSEs for non-sampled areas of SVPG and PCAR methods over those of SCPG method.}
\medskip
\begin{tabular}{lrrrrrrr}
\hline
\multicolumn{1}{r}{$n_i$} & 5 & 10 & 20 & 30 & 40 & 60 & 100\\
\hline
SVPG & 1.162 & 0.830 & 0.780 & 0.839 & 0.980 & 0.622 & 0.979\\
PCAR & 1.149 & 0.751 & 0.906 & 0.952 & 1.031 & 0.694 & 0.982\\
\hline
\end{tabular}
\label{tab:PG-nonsample}
\end{center}
\end{table}

\begin{table}[htbp]
\begin{center}
\caption{The ratios of the MSEs for non-sampled areas of SVBB and LCAR methods over those of SCBB method.}
\medskip
\begin{tabular}{lrrrrrrr}
\hline
\multicolumn{1}{r}{$n_i$} & 5 & 10 & 20 & 30 & 40 & 60 & 100\\
\hline
SVBB & 1.081 & 0.961 & 0.932 & 0.893 & 0.976 & 0.888 & 1.018\\
LCAR & 1.098 & 0.954 & 1.002 & 0.954 & 0.998 & 0.939 & 1.036\\
\hline
\end{tabular}
\label{tab:BB-nonsample}
\end{center}
\end{table}

\begin{table}[htbp]
\begin{center}
\caption{The relative bias and coefficient of variations of two types of MSE estimators (bootstrap estimator and naive estimator) based on Poisson--Gamma model. The values are averaged over the groups within the same $n_i$ values. RB and RBN denote relative bias of bootstrap estimator and naive estimator, respectively, and CV and CVN denote coefficient of variations of bootstrap estimator and naive estimator, respectively.}
\medskip
\begin{tabular}{llrrrrr}
\hline
& \multicolumn{1}{r}{$n_i$} & 10 & 15 & 20 & 25 & 30\\
\hline
\multirow{4}{*}{$m=30$} & RB & 0.091 & 0.073 & 0.096 & 0.035 & 0.039\\
& RBN & -0.206 & -0.266 & -0.189 & -0.244 & -0.225\\
& CV & 0.581 & 0.475 & 0.462 & 0.401 & 0.374\\
& CVN & 0.606 & 0.569 & 0.487 & 0.503 & 0.484\\
\hline
\multirow{4}{*}{$m=50$} & RB & 0.134 & 0.009 & 0.085 & -0.053 & 0.09\\
& RBN & -0.086 & -0.277 & -0.218 & -0.281 & -0.171\\
& CV & 0.570 & 0.476 & 0.459 & 0.407 & 0.404\\
& CVN & 0.521 & 0.523 & 0.474 & 0.487 & 0.438\\
\hline
\multirow{4}{*}{$m=80$} & RB & 0.153 & 0.067 & 0.056 & 0.075 & -0.030\\
& RBN & 0.058 & -0.081 & -0.120 & -0.051 & -0.167\\
& CV & 0.607 & 0.508 & 0.443 & 0.408 & 0.347\\
& CVN & 0.560 & 0.485 & 0.419 & 0.395 & 0.392\\
\hline
\end{tabular}
\label{tab:PG-mseestimate}
\end{center}
\end{table}

\begin{table}[htbp]
\begin{center}
\caption{The relative bias and coefficient of variations of two types of MSE estimators (bootstrap estimator and naive estimator) based on binomial--beta model. The values are averaged over the groups within the same $n_i$ values. RB and RBN denote relative bias of bootstrap estimator and naive estimator, respectively, and CV and CVN denote coefficient of variations of bootstrap estimator and naive estimator, respectively.}
\medskip
\begin{tabular}{llrrrrr}
\hline
& \multicolumn{1}{r}{$n_i$} & 10 & 15 & 20 & 25 & 30\\
\hline
\multirow{4}{*}{$m=30$} & RB & 0.133 & -0.019 & -0.034 & -0.026 & -0.049\\
& RBN & -0.267 & -0.400 & -0.385 & -0.372 & -0.409\\
& CV & 0.604 & 0.429 & 0.430 & 0.387 & 0.340\\
& CVN & 0.580 & 0.566 & 0.545 & 0.534 & 0.547\\
\hline
\multirow{4}{*}{$m=50$} & RB & 0.157 & 0.088 & 0.011 & 0.004 & 0.030\\
& RBN & -0.111 & -0.210 & -0.282 & -0.254 & -0.230\\
& CV & 0.593 & 0.480 & 0.415 & 0.375 & 0.376\\
& CVN & 0.505 & 0.485 & 0.476 & 0.440 & 0.435\\
\hline
\multirow{4}{*}{$m=80$} & RB & 0.113 & 0.083 & -0.006 & 0.008 & -0.011\\
& RBN & -0.054 & -0.117 & -0.226 & -0.163 & -0.200\\
& CV & 0.615 & 0.504 & 0.410 & 0.375 & 0.337\\
& CVN & 0.537 & 0.467 & 0.440 & 0.396 & 0.393\\
\hline
\end{tabular}
\label{tab:BB-mseestimate}
\end{center}
\end{table}

\begin{table}[htbp]
\caption{
Quantiles of hyperparameter estimates in the spatially varying (SV) model and point estimates in the spatially constant (SC) model, and $p$-values for testing spatial variation using Spanish poverty rate data.
\label{tab:BB-pov}
}
\begin{center}
\medskip
\begin{tabular}{crrrrrrc}
\hline
& \multicolumn{5}{c}{SV} & SC & Spatial variation\\
 & $0\%$ & $25\%$ & $50\%$ & $75\%$ & $100\%$ & Estimate & $p$-value\\
\hline
$\beh_1$ & -7.96 & -3.34 & -2.28 & -1.03 & 1.11 & -2.70 & 0.052\\
$\beh_2$ & -3.47 & 2.13 & 3.50 & 5.26 & 10.65 & 3.85  & 0.065\\
$\beh_3$ & -4.54 & -2.16 & -1.69 & -0.90 & 3.22 & -1.19  & 0.113\\
$\nuh$ & 42.33 & 44.12 & 48.06 & 51.59 & 103.06 & 46.32  & 0.762\\
\hline
\end{tabular}
\end{center}
\end{table}

\end{document}